Mapping nanoscale thermal transport in liquid environment - immersion scanning thermal microscopy
**Peter D Tovee and Oleg V Kolosov[1]**
Physics Department, Lancaster University, Lancaster, LA1 3BE, UK.

[1]Author to whom any correspondence should be addressed

E-mail: o.kolosov@lancaster.ac.uk

**Abstract.**

Nanoscale heat transport is of increasing importance as it often defines performance of modern processors and thermoelectric nanomaterials, and affects functioning of chemical and biosensors. Scanning Thermal Microscopy (SThM) is the leading tool for nanoscale mapping of thermal properties, but it is often negatively affected by unstable tip-surface thermal contacts. While operating SThM in liquid environment may allow unimpeded thermal contact and open new application areas, it was so far regarded impossible due to increased heat dissipation into liquid, and perceived non-local thermal interaction between the probe and the sample. Nevertheless, in this paper we show for the first time that such liquid immersion SThM (iSThM) is fully feasible and has sufficient thermal contrast to detect thermal conductivity variations in the few tens of nm thick graphite nanoflake and metal-polymer nanostructured interconnects. Its spatial resolution was on the order of 50 nm, equal to the resolution of the same probe in the standard air environment. These results confirm localized thermal sensing in iSThM and, coupled with absence of tip snap-in due to elimination of capillary forces, suggest the possibility for true non-contact nanoscale thermal mapping in liquids, including thermal phenomena in energy storage devices, catalysts and biosystems.

PACS:

## 1. Introduction.

Since its invention, scanning probe microscopy (SPM) [1, 2] became an indispensable tool of modern nanotechnology. One of major advantages of SPM is its ability to sense diverse physical and chemical properties of nanostructures with nanometer resolution [3-5] including operation in various environments [6]. In particular, nanoscale temperature measurements [7], heat generation and nanoscale heat propagation are of increasing importance due to continuous decrease of size of semiconductor devices with concurrent increase of processing power [8-10]. While active Scanning Thermal Microscopy (SThM) [11-22] that uses self-heated thermal sensors in contact with the studied sample allows nanoscale mapping of thermal properties, a weak thermal coupling between sensor and the sample is one of major SThM problems [16]. Furthermore, this coupling fluctuates while scanning and is also affected by sample geometry, overall significantly reducing SThM performance in nanoscale thermal measurements.

It would be very tempting to use liquid immersion in SThM to improve both the thermal contact between the probe tip and the sample as well as contact uniformity. The approach would be somewhat similar to one used in ultrasonic imaging where dedicated gel is used to achieve better acoustic coupling [23] or optical microscopy where immersion reduces light reflection, refraction and scattering at the interfaces [24]. While the role of liquid film between the SThM tip apex is known to be essential [16, 17, 20, 21] up to now there were no reports on heated nanoscale resolution SThM imaging while probe fully immersed in the liquid. So far SThM operation both in air and vacuum was demonstrated, with vacuum helping to eliminate heat dissipation through air [14, 19, 25]. The only reported so far in-liquid nanoscale thermal measurements used passive fluorescence thermometry [26], that while having potential for sub-micrometer spatial resolution [27] is not capable of measurements of local thermal transport or thermal conductivities as it relies on the externally created heat flux. If feasible, SThM liquid immersion would improve and stabilize thermal contact between the tip and the surface, compared to in-air or vacuum environment. Also, due to efficient heat transfer through liquid, it might be possible to perform truly non-contact scanning with the tip-surface separated by few nm gap, while retaining both nanoscale resolution and thermal sensitivity. Such in-liquid SThM, would be of extreme interest for biotechnology, where it will allow handling of delicate biological samples [28] and in exploration of energy generation and dissipation in rechargeable batteries, fuel cells and liquid phase catalysts.

Unfortunately, until now fully immersed operation of SThM has been considered all but impossible due to the potentially overwhelming direct heat dissipation from the heated sensor into the surrounding liquid, that was perceived to result in the non-local thermal sensing and degradation of lateral resolution [16]. To our knowledge, there is so far no publication reporting such approach and measurements. Notwithstanding this rationale, in this paper we show that such immersion SThM, or iSThM, can be successfully realized using certain design of a SThM probe that is fully immersed in a liquid [29]. We found, surprisingly, that iSThM performance would not be qualitatively different from the in-air or in-vacuum environment operation [19, 20, 25]. We then apply iSThM to explore the heat transport in polymer-metal ultra large scale integration interconnects [30], and in few tens of nm thick graphite nanoflake [31] on Si substrate. In these nanostructures we observe 50 nm lateral resolution to local thermal conductance that confirmed a local nature of iSThM thermal contrast, with experimental finding supported by the numerical simulations that allowed to understand the underlying nature and imaging mechanisms of a new approach.

## 2. Methods and materials

*2.1 Finite elements modeling of SThM probes in air and liquid environment.*

In order to explore the feasibility of iSThM and to find optimal experimental approach, we first simulated behavior of various SThM probes in air and liquid environments. The simulations were based on commercial finite elements (FE) (COMSOL Multiphysics) approach in the 3D probe geometry. We used COMSOL AC/DC module for modeling the current flow, Joule heating of the probe and a probe resistance that reflected probe temperature, as well as thermal module for modeling of heat transport [29]. Given that mean-free-path length of heat carriers (phonons in dielectrics and electrons in metals) in relevant materials in the model were generally on the order of 10-25 nm, and the minimal characteristic dimensions of the probe geometrical elements were 50 nm or larger, a diffusive approximation and Fourier heat transport equations were deemed to be adequate [19].

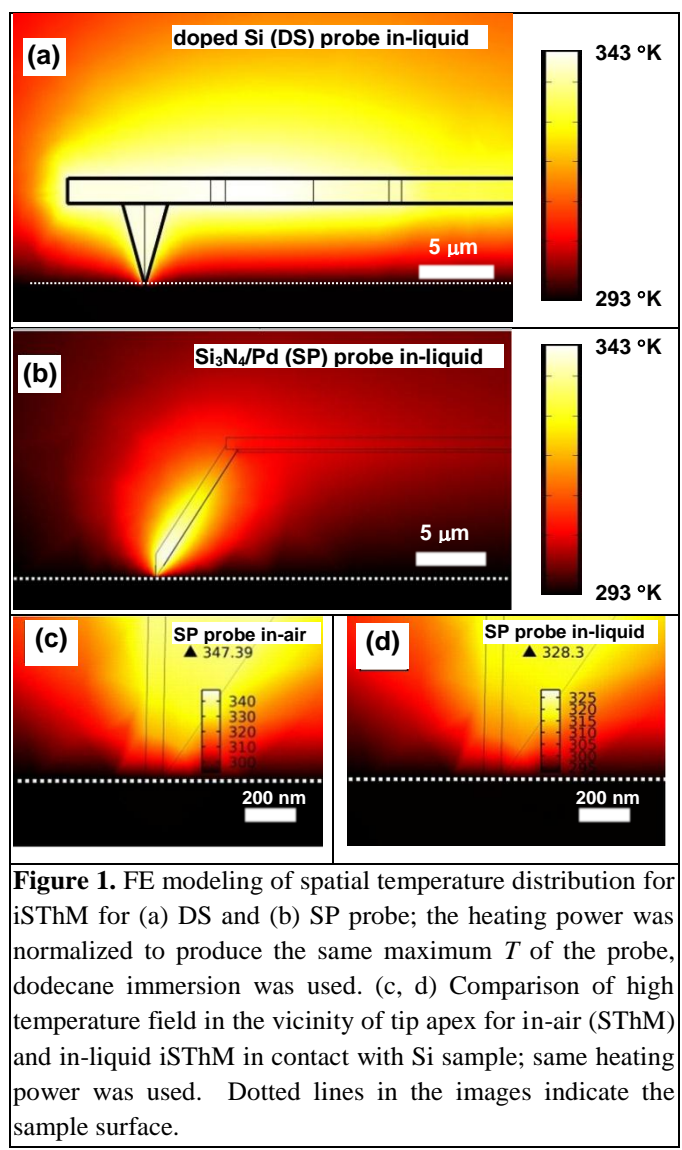

**Figure 1.** FE modeling of spatial temperature distribution for iSThM for (a) DS and (b) SP probe; the heating power was normalized to produce the same maximum $T$ of the probe, dodecane immersion was used. (c, d) Comparison of high temperature field in the vicinity of tip apex for in-air (SThM) and in-liquid iSThM in contact with Si sample; same heating power was used. Dotted lines in the images indicate the sample surface.

At first glance, the direct heat dissipation from SThM probe to the surrounding liquid might result in the complete loss of sensitivity to local thermal properties of the sample and lead to non-localized thermal response [16, 20]. In fact, FE modeling of a widely used doped Si probe (DS) [13, 29] (AN-200, TA Instruments) in dodecane (this liquid was selected for its low-volatility, moderate thermal conductivity and a non-corrosiveness) supported such general belief showing that the heat plum is concentrated near the heater and far from the sample (Figure 1(a)). Such heat distribution would indeed significantly decrease sample related thermal signal and spatial resolution. Surprisingly, at the same time, FE analysis of another widely used design of $Si_3N_4$ probe with Pd resistive element at the end of the probe tip (SP probe) (Kelvin Nanotechnologies) [29, 32] showed a distinctive narrow heat plum concentrated near the end of the probe (Figure 1(b). The absolute value of the SP probe temperature with respect to the environment was only slightly lower in liquid - 35 degrees vs 54 degrees in air, corresponding to only 35% loss of thermal signal (see temperature scale in Figure 1 (c,d). More significantly, a temperature field near the probe apex shows almost identical pattern of the spatial distribution of temperature for SP probe in- air (Figure 1(c)) and in liquid (Figure 1(d)). These finding strongly suggested that SP should be a probe of choice for iSThM experimental trials.

*2.2. Experimental setup of SThM and thermal measurements.*

Our iSThM experiments used a general purpose SPM (Bruker Multi-Mode, Nanoscope III) with "half-moon" SThM probes holder (Anasys Instruments) that was modified for using thermal probe in liquids as shown in Figure 2. A special PTFE cup was used to contain the liquid as well as to keep the sample, thermal cantilevers and connecting leads immersed in liquid. A glass window was used to create a flat refraction interface for SPM laser beam that monitors cantilever deflection. A nitrile rubber thin sheet was used underneath the liquid holder to protect the scanner from accidental liquid spilling.

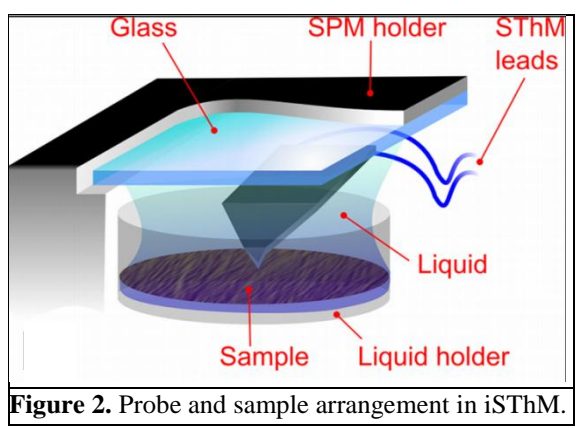

**Figure 2.** Probe and sample arrangement in iSThM.

The SP thermal probe was calibrated on a Peltier variable temperature plate (Echo Therm IC20, Torrey Pines Scientific) at temperatures ranging from room temperature to 80°C by measuring probe electrical resistance as a function of the applied voltage and ambient temperature. Two types of resistance measurements were used – in the first the probe formed a part of a DC voltage divider in series with the fixed resistor, this method was found to be highly accurate but slow and used predominantly during the calibration. The second method combined low AC voltage signal at 91 kHz frequency that was used for resistance measurements, and higher DC offset providing probe self-heating. In this method the probe formed a part of a balanced Maxwell electric bridge, with its output measured via the lock-in amplifier (SRS-830, Stanford Research Systems). In second mode the resulting signal rapidly followed the resistance changes but had less absolute accuracy, this mode was mainly used for real-time SThM imaging. The AC and DC voltage excitation for all measurements was provided by the precision function generator (Model 3390, Keithley Instruments). Both approaches indicated a linear dependence of the probe resistance on its temperature that was expected as the metallic resistive sensing element was used. As the tip is brought in contact with the surface during SThM imaging, some heat starts to flow into the sample cooling the tip and, consequently, changing resistance of the sensor. Using a Maxwell bridge and combined AC-DC excitation [29] the well-defined power can be applied to the sensor generating Joule heat, with the tip temperature measured simultaneously. As the tip is scanned across the sample, monitoring these values allows to create a thermal image with the lower temperature of the probe corresponding to the higher local heat flux into the sample, therefore allowing direct evaluation of its local thermal conductivity [16, 19, 33].

*2.3 Sample preparation.*

Two samples used in this study were selected for two interlinked purposes. First we targeted to prove experimentally a feasibility of iSThM, its sensitivity to local thermal conductivity, and thermal spatial resolution while imaging diverse materials, such as metals, polymers, and semiconductors. Secondly, if iSThM approach is successful, we were interested in experimental exploration of differences of nanoscale thermal transport in air and in the liquid environment when nature of thermal contact between the material and surrounding liquid immersion may modify such transport. First sample was a metal-polymer nanostructure, where Al damascene embedded layer [30] was enclosed in benzocyclobutene (BCB) low-k dielectric, this sample was used to represent morphology and materials of the ultra large scale integration interconnects (ULSI). Heat transport in ULSI nanostructures is of significant concern as metallization tracks in a real device carry significant amount of current and their overheating would increase the electro-migration and adversely affect the stability of the semiconductor device [30]. High resolution lithography used in the sample preparation provided a well-defined boundary between metal and BCB, while the chemo-mechanical polishing provided relatively flat surfaces with moderate topographical contrast. The second sample was in-house exfoliated a few layer graphene/graphite nanoflakes with thicknesses up to few tens of nm deposited on Si wafer with 300 nm thermal oxide layer on the top. The ULSI interconnects were cleaned by 10 min sonication consecutively in acetone, ethanol, and DI water, with final plasma cleaning performed in $O_2$/Ar plasma for 5 minutes. Such stringent procedure removes any residual organic contaminants on the surface. The Si wafer prior to the deposition of nanoflakes was cleaned using the same procedure, and nanoflake was then mechanically exfoliated from Kish graphite lumps using pressure sensitive tape, with the final exfoliation step performed by cross-linked polymer gel (Gel-Pak, USA) that helped to minimize tape residue transfer [19].

3. Results and discussion

*3.1 Feasibility of iSThM: sensing thermal conductivity with nanoscale resolution and local nature of iSThM response.*

The feasibility of liquid environment iSThM operation was first tested on ULSI interconnect structures described above. The measurements were performed by the probe first operated in-air and then in-the-liquid dodecane environment, allowing making relevant and definitive comparisons as exactly the same probe was used. Topography images in both environments shown in Figure 3(a,b) reveal similar features - a protruding central Al metallization lead with few voids (that are usually located at Al grain boundaries [34]) embedded in the BCB polymer matrix. The in-air SThM image in Figure 3(c) obtained at a constant Joule heating power applied to the sensor, clearly shows that the Al interconnects produce better heat dissipation (darker image contrast reflects the lower temperature of the probe due to increased heat flow from the probe). This is consistent with the higher thermal conductivity for Al, $k_{Al}$ = 200 $Wm^{-1}K^{-1}$ compared to thermal conductivity of BCB, $k_{BCB}$ = 29 $Wm^{-1}K^{-1}$. The absolute topographical height (~ 80 nm) seemed to provide no direct influence for in-air thermal image, at the same time the "rim" of Al lead looked brighter. One can expect certain increase of heat transport at the Al edge due to increased contact area with the side of SThM tip, and hence darker SThM contrast, but this was not observed. Most likely the nanoscale roughness of the side edge created a barrier to the heat transport resulting in it imaged as "hotter" - brighter areas in in-air SThM images.

iSThM thermal image in Figure 3(d) shows qualitatively same features – darker Al metallization layer with better heat dissipation surrounded by the brighter, lower heat conductance BCB matrix. As for in-air SThM thermal image, the absolute topographical height of the Al layer (~ 80 nm) provided no direct influence on the iSThM thermal image, confirming the local nature of the probing. The lateral resolution of iSThM was similar to the resolution of in-air SThM, as seen from the comparison of line profiles given in Figure 3(e).

This was further supported by FE simulation of the same profiles in Figure 3(d) where the thermal signal across the interface between Al and BCB polymer shows excellent qualitative correspondence with experimentally obtained SThM and iSThM profiles. In both modes, the thermal signal on the Al – BCB polymer boundary shows appreciable change over approximately 50 nm distance, in good correlation with the experimental results. The experimentally measured signal-to-noise ratio of iSThM thermal image was lower compared to in-air SThM, that was most likely linked with the additional heat dissipation into the liquid adding to the common for SThM and iSThM modes heat dissipation to the cantilever base. Nevertheless, the thermal signal was fully sufficient for nanoscale mapping of thermal conductivity in ULSI nanostructure in iSThM. While more definitive comparison of the lateral resolution in iSThM and SThM is given later in the paper using a well-defined edge of graphite nanoflake, data in Figure 3 clearly indicate that iSThM operating in liquid environment is feasible and capable of nanoscale mapping of local thermal conductivities.

The interesting observation was that in contrast to in-air SThM, the voids in Al (observed in both topographical images) did not produce significant increase in thermal response in iSThM suggesting that the immersion liquid secured direct thermal contact of the probe and the studied material. Also a darker edge around Al lead that indicated higher heat transport to the side of the lead was observed without lighter "rim" in iSThM. These observations suggest that iSThM has potential of eliminating detrimental effects of surface roughness, and can reflect more directly the intrinsic heat transport in nanostructures.

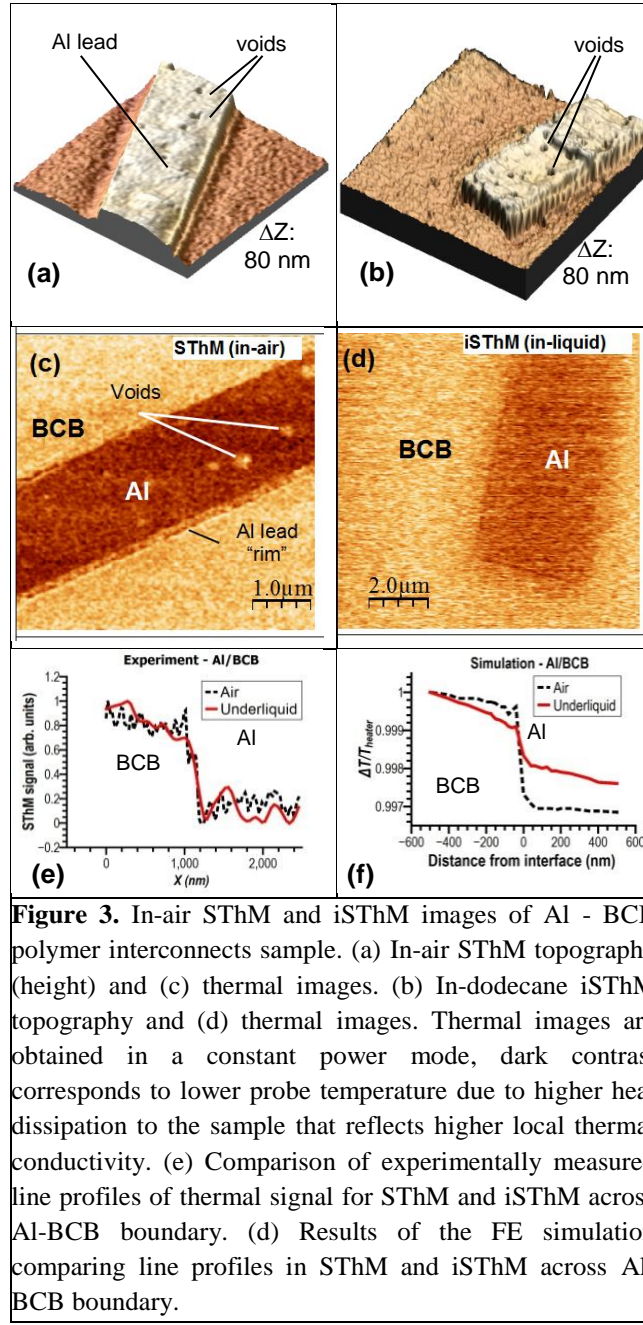

**Figure 3.** In-air SThM and iSThM images of Al - BCB polymer interconnects sample. (a) In-air SThM topography (height) and (c) thermal images. (b) In-dodecane iSThM topography and (d) thermal images. Thermal images are obtained in a constant power mode, dark contrast corresponds to lower probe temperature due to higher heat dissipation to the sample that reflects higher local thermal conductivity. (e) Comparison of experimentally measured line profiles of thermal signal for SThM and iSThM across Al-BCB boundary. (d) Results of the FE simulation comparing line profiles in SThM and iSThM across Al-BCB boundary.

*3.2 iSThM mapping of thermal transport in thermally anisotropic graphite nanoflake.*

The iSThM was then applied to investigate thermal conductivity of graphite nanoflakes of approximately 40 nm thickness on Si substrate (Figure 4). Thermal properties of a few layer graphene and graphite exhibit extremely anisotropic thermal conductivity that depends on the direction of the heat flux. Ratio of thermal conductivity parallel and perpendicular to graphene planes [35] can reach three orders of magnitude ($k_{\parallel\text{-}Gr\text{-}plane}$ = 2000 Wm$^{-1}$K$^{-1}$ $k_{\perp\text{-}Gr\text{-}plane}$ = 2 Wm$^{-1}$K$^{-1}$) with thermal conductivity in graphene plane exceeding one of Si ($k_{Si}$ = 130 Wm$^{-1}$K$^{-1}$), whereas conductivity normal to graphene planes being two orders of magnitude lower that thermal conductivity of Si. Such aspects may play some role in the heat dissipation in rechargeable batteries, supercapacitors and fuel cells where graphene and graphite is used in liquid environment [36] with iSThM offering a feasible approach for such studies.

Thermal images in Figure 4(c,d) and corresponding line profiles (Figure 4(e,f) show that for in-air SThM the heat transport is slightly increased in the area of the flake, and the side of the flake, with clear decrease of heat transport at the flake "rim" (tip position *iii*, Figure 4(g,h). At the same time, iSThM thermal image seem to be less affected by the "rim" contrast, presumably, due to better thermal link between the tip and the sample that is liquid mediated and less affected by the local surface corrugations. Moreover, the heat transport in the area of the flake (area *iv*, as illustrated in the Figure 5(g,h) is slightly below one of Si. That can be linked with the fact that in-liquid a higher proportion of the heat flux is normal to graphene layers along the direction of the lower thermal conductivity ($k_{\perp\text{-}Gr\text{-}plane} = 2$ Wm$^{-1}$K$^{-1}$ compared to $k_{Si} = 130$ Wm$^{-1}$K$^{-1}$) [37] or with different interfacial Kapitza resistance [38]. The side surface of the flake provides a better heat dissipation (tip position *ii*) for both in-air and iSThM due to high in-plane conductivity of graphene layers ($k_{\parallel\text{-}Gr\text{-}plane} = 2000$ Wm$^{-1}$K$^{-1}$). The FE simulation of the profile across nanoflake-substrate boundary in Figure 4(i) show slightly increased heat transport for in-air SThM in the flake area compared to Si substrate, whereas heat transport for iSThM being reduced in the flake area. The sign of the heat dissipation difference matches well with the one experimentally observed in Figure 4(e,f). Moreover, the simulation shows that the spatial variation of the heat dissipation due to presence of the flake "rim" are reduced in iSThM in good correspondence with experimental data.

Finally, it is to be noted that the full width of half maximum (FWHM) of the flake edge for iSThM was 56 nm compared with 78 nm for in-air SThM obtained with the same probe (Figure 4(e,f), confirming the local nature of iSThM thermal response and indicating at least similar lateral resolution for comparable probes and the samples.

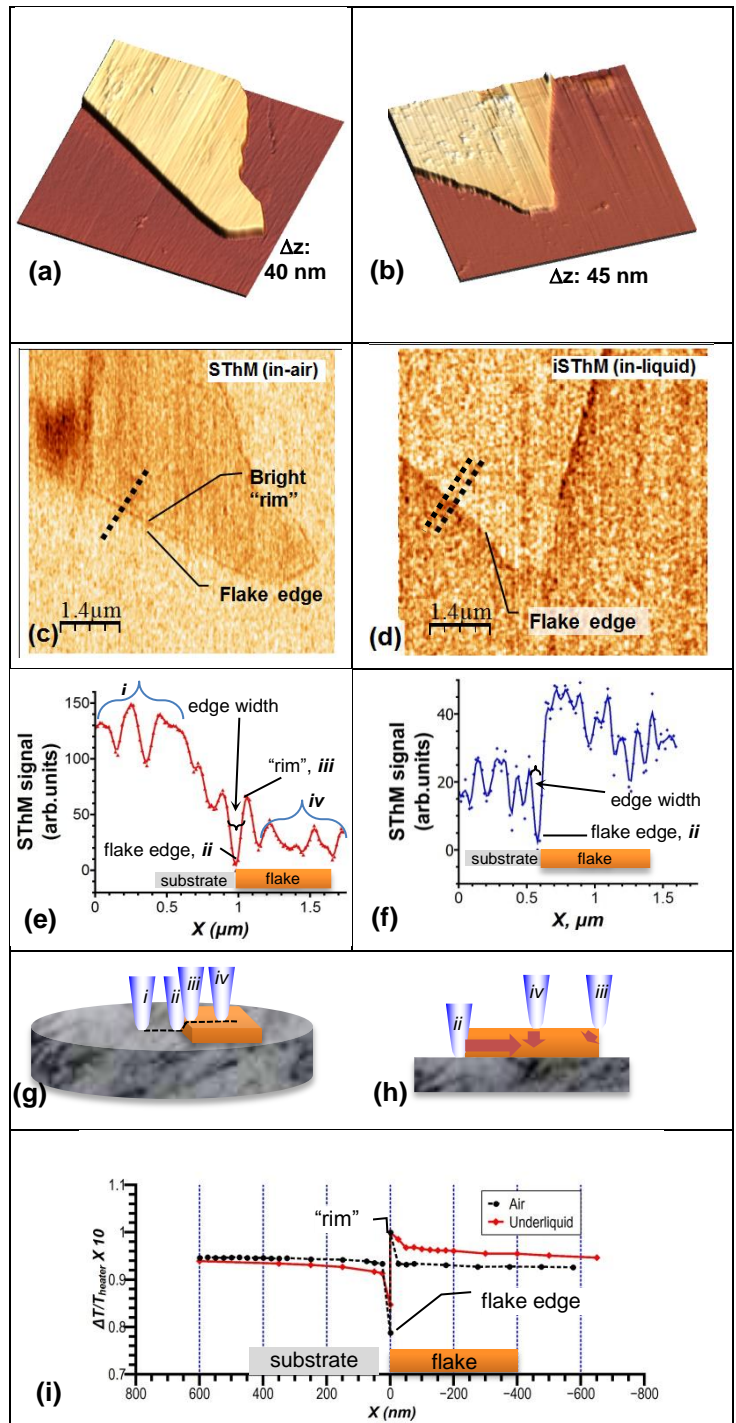

**Figure 4.** SThM and iSThM nanoscale mapping of thermal conductivity in graphite nanoflakes. (a) In-air topography and (c) corresponding thermal image of 40 ± 5 nm thick nanoflake. iSThM (b) topography and (d) thermal image of a similar flake with 45 ± 5 nm thickness. (g, h) Schematic illustration of different regions in the images with SthM tip positioned on: *i* - the substrate, *ii* – touching the edge of the flake (thermal contact area increased), *iii* – on the "rim" of the flake (contact area decreased), *iv* - on the top of the flake. (i) Simulated SThM and iSThM profiles across the edge of graphite nanoflake.

*3.3 Force contact phenomena in iSThM.*

As we observed above, iSthM provides efficient through-the-liquid thermal coupling of the probe and the sample that eliminates some effects of the surface roughness and voids, e.g. in ULSI in Figure 2. In principle, that would allow the true non-contact thermal imaging with the tip hovering at few nm above the studied surface, and enable study of delicate samples such as biological materials, porous and suspended nanostructures. We have performed FE simulations that indicated that 50 nm wide tip "hovering" above sample surface at 25 nm distance would retain 80% of the iSThM in-contact thermal signal, whereas in-air SThM will retain less than half of such signal.

In order to analyze the non-contact performance of iSThM, we captured simultaneously thermal signal and mechanical deflection of SThM sensor (proportional to force between the sample and SThM tip) as the tip approached the sample (Figure 5) similar for in-air measurements in [20]. In either environment as the tip approaches the surface, the thermal signal notably decreases, as the heat conducted to the sample via either air [20] or liquid media cools the probe. For in-air SThM (Figure 5(a,b) there is a well-defined "snap-in" of the tip towards the surface that is linked with the well-documented formation of meniscus [20, 37] that also leads to the step-wise increase of the pull-in force helping to establish the solid-solid contact. Such behavior is typical for the relatively compliant cantilever with the sharp tips like the one used in our study. Once the capillary meniscus and solid-solid contact is fully established, the thermal signal remains mainly constant even as the contact force increases. When the sample is retracted, the combined adhesive and meniscus forces lead to even stronger "snap-off" that results in the jump-off of the cantilever and similar changes in the thermal response.

In the liquid environment of iSThM the situation was quite different – no clear "snap-in" or "snap-out" force jumps were observed (Figure 5(c) as it would be expected due to the absence of capillary forces [39]. Also, no jump of heat dissipation (Figure 5(d) on the solid-solid contact was observed as it would be for in-air or in-vacuum environments [25]. The absence of "thermal" jump in combination with the spatial resolution demonstrated in Figures 3(e) and 4(f) suggest that iSThM sensing of local thermal conductivity of the sample is almost fully liquid mediated. The observed absence of the capillary "snap-in" and "snap-out" in iSThM coupled with demonstrated in this paper similar to in-air SThM 50 nm lateral resolution to thermal properties, would make it possible to realize true non-contact nanoscale thermal imaging by "hovering" iSThM tip in the few to 10 nm above the sample, maintaining the distance via *e.g.* well-known shear force feedback [40] that would be easily realized due to non-zero viscosity of coupling liquid.

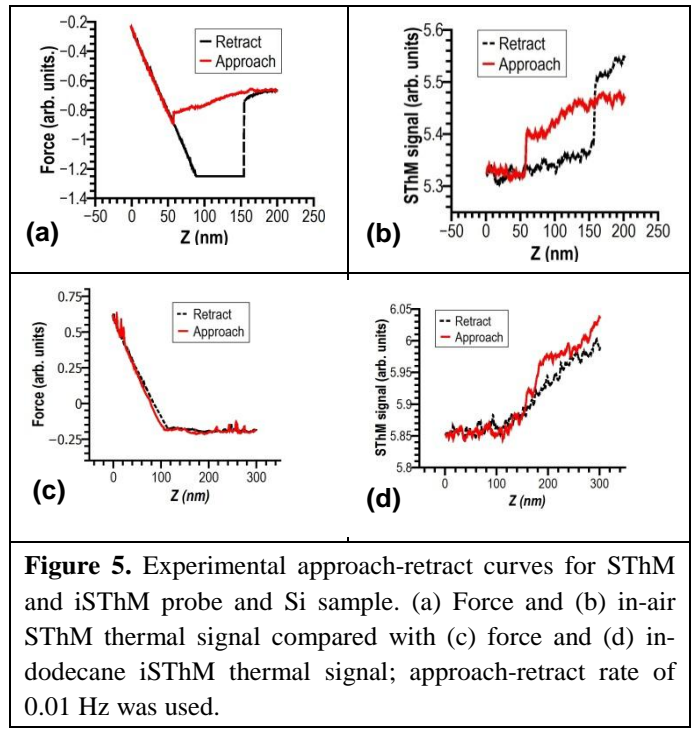

**Figure 5.** Experimental approach-retract curves for SThM and iSThM probe and Si sample. (a) Force and (b) in-air SThM thermal signal compared with (c) force and (d) in-dodecane iSThM thermal signal; approach-retract rate of 0.01 Hz was used.

## 4. Conclusions

This study demonstrates for the first time that active heating nanoscale SThM thermal probing in a fully immersed liquid environment (iSThM) is feasible. The iSThM was shown to be sensitive to the local thermal conductivity of materials ranging from polymers (BCB) and semiconductors (Si) to metals (Al) and graphite nanoflakes. The effective lateral resolution on the order of 50 nm was demonstrated using a widely used microfabricated resistive probe design operating in iSThM mode, and shown to be very similar to the lateral resolution of the same probe operated in air. Using FE modeling we found that the probe design may play crucial role for iSThM performance, suggesting that dedicated optimization of probe for iSThM, e.g. changing the geometry of the heater and the probe apex, can significantly improve iSThM resolution and performance. The iSThM thermal response was shown to be in good qualitative agreement with numerical simulations, that supported experimentally observed iSThM spatial resolution and sensitivity. We confirmed that the heat flux between the probe and the sample that defines the nanoscale spatial resolution of iSThM is predominantly conducted through liquid rather solid-solid contact, that may help to eliminate one of major uncertainties in the nanoscale mapping of heat transport via active SThM – instability of tip-surface thermal contact. Moreover, such liquid mediated iSThM should make it possible to realize true non-contact nanoscale thermal imaging by "hovering" iSThM tip in the vicinity of the sample. We believe that iSThM can open new possibilities for exploration of heat generated in biological systems [41, 42], and help to investigate nanoscale thermal phenomena in power batteries, fuel cells, and nanoscale catalysts.

**Acknowledgment**

Authors acknowledge input of Manuel Pumarol for advice, scientific discussions and support related to the variety of aspects of SThM operation. We appreciate help of Riccardo Mazzocco with some of illustrations and grateful to Bob Geer for providing ULSI samples, and Bob Jones for SEM analysis of the probes. O.V.K. acknowledges support from the EPSRC grants EP/G015570/1, EP/K023373/1, EPSRC-NSF grant EP/G06556X/1 and EU FP7 GRENADA and FUNPROBE grants. Authors acknowledge use of WSxM and Gwiddion for analysis of SPM images.